\documentclass[aps,twocolumn,groupedaddress]{revtex4}
\usepackage{amsmath,amssymb}

\begin{document}
% You should use BibTeX and revtex.bst for references
\bibliographystyle{apsrev}

%\preprint{}

\title{$T^\prime=T$,
$S^\prime=\gamma S$,
${\cal P}^\prime = \gamma^2 {\cal P}$,
${\cal V}^\prime ={\cal V}/\gamma$,
and $U^\prime = \gamma U$}

% Optional argument for running titles on pages %\title[]{}


% \affiliation command. The \affiliation command MUST follow the

\author{Ryszard Zygad\l{}o }
%\email[]{Your e-mail address} %\homepage[]{Your web page}
%\thanks{}
\affiliation{Marian Smoluchowski
Institute of Physics, Jagiellonian University, Reymonta 4, PL--30059 Krak\'ow,
Poland}


\date{\today}

\begin{abstract}
It is shown that the  Lorentz invariant temperature
$T^\prime=T$ and
entropy $S^\prime=\gamma S$,
pressure ${\cal P}^\prime = \gamma^2 {\cal P}$,
volume ${\cal V}^\prime ={\cal V}/\gamma$,
and energy $U^\prime = \gamma U$,
where $\gamma = (1-V^2/c^2)^{-1/2}$, are the correct
transformational properties of the thermodynamical
quantities of the relativistic ideal gas.
\end{abstract}

%\maketitle must follow title, authors, abstract and PACS
\maketitle

%\section{}
%\label{}
%\subsection{}
%\subsubsection{}

%\paragraph*{1.}
The old problem of transformational
properties, see Ref.~[1] for details,
of the thermodynamical quantities,
which enter to the basic equation,
\begin{equation}
dU^\prime = T^\prime dS^\prime - {\cal P}^\prime d{\cal V}^\prime,
\end{equation}
in the moving (along $x$-axis with the constant velocity $V$)
reference frame, is considered for the ideal relativistic gas.
The basic thermodynamical definitions are used and new results
are obtained.

The description (1) --
{\it thermodynamics}
-- requires the following:

(i) all three terms of Eq.~(1) transform according to the same rule;

(ii) the volume of the box transforms like ${\cal V}^\prime ={\cal V}/\gamma$,
where $\gamma = (1-V^2/c^2)^{-1/2}$;

(iii) the energy transforms like
\begin{equation}
U^\prime = \sum_i \langle \varepsilon^\prime_i \rangle
= \sum_i \langle \gamma [\varepsilon_i - (p_x)_i V] \rangle
=\sum_i \gamma \langle \varepsilon_i \rangle = \gamma U,
\end{equation}
where
$\langle (p_x)_i \rangle =0$ in the common resting
reference frame of a system (as a whole) and thermal bath;

(iv) the entropy is known (as well as the other thermodynamical quantities)
{\it explicitly} in the resting reference frame from Gibbs--Boltzmann equilibrium distribution,
particularly
\begin{equation}
dS = \phi_1(T)dT + \phi_2({\cal V}) d{\cal V},
\end{equation}
where $\phi_1 [= Tc_{\cal V}]$ and $\phi_2$  is a  (known)
{\it nonhomogeneous} and {\it homogeneous}
(of the order $-1$, which is going to appear {\it crucial})
function
of the single argument, respectively.

The rule for the pressure,
\begin{equation}
{\cal P}^\prime = \gamma^2 {\cal P},
\end{equation}
follows as a {\it necessary condition} for the
selfconsistence of thermodynamics.
The same result (4) is
independently obtained from the mechanical
consideration. In fact, considering
the point particles (i.e., ideal gas)
we can neglect the collisions between particles. Then each particle
undergoes the successive collisions with the certain wall of the box
($L\times L \times L ={\cal V}$) with
certain constant period
$\tau_u = 2L\varepsilon/p_u $ , $u=x,\, y, \,z$,
respectively in each direction.
In the moving reference frame all $\tau_u$
change according to the Lorentz dilatation
$\tau_u^\prime = \tau_u/\gamma$.
The momentum transfer associated with the single collision in
$y$ and $z$ direction is the same $2p_y^\prime=2p_y$,
$2p_z^\prime=2p_z$ in the moving reference frame and the areas
change according to $\sigma_y^\prime = \sigma_y/\gamma$,
$\sigma_z^\prime = \sigma_z/\gamma$. The energy--momentum transfer
in $x$-direction $(0,\,2p_x,\,0,\,0)$ corresponds to
$(-2V\gamma p_x,\,2\gamma p_x,\,0,\,0)$ in the moving reference frame
and $\sigma_x^\prime = \sigma_x$. Consequently the each
ratio $2p_u/\tau_u \sigma_u$ change by the factor $\gamma^2$ and
thus ${\cal P}^\prime = \gamma^2 {\cal P}$.

Comparing
\begin{equation}
dU^\prime = T^\prime dS^\prime -{\cal P}^\prime d{\cal V}^\prime
=\gamma (TdS -{\cal P} d{\cal V}) =\gamma dU
\end{equation}
and using the pressure transformation formula (4) one obtains
\begin{equation}
(T^\prime/\gamma T) dS^\prime(T^\prime, {\cal V}/\gamma) = dS(T,{\cal V}).
\end{equation}
Consequently, see Eq.~(3),
$(T^\prime/\gamma T) \partial S^\prime/\partial T^\prime \equiv
(T^\prime/\gamma T) \psi_1 (T^\prime)$ cannot depend on
${\cal V}^\prime$. Because
$0 = \partial \psi_1/ \partial{\cal V}^\prime =
\partial^2 S^\prime/ \partial{\cal V}^\prime \partial T^\prime$
one obtains
$\partial S^\prime/\partial
{\cal V}^\prime \equiv \psi_2 ({\cal V}^\prime)$ depending only on
${\cal V}^\prime$. The equality
\begin{equation}
(T^\prime/\gamma^2 T) \psi_2 ({\cal V}/\gamma) = \phi_2({\cal V})
\end{equation}
can only be satisfied ({\it identically} with respect to ${\cal V}$) if
\begin{equation}
T^\prime = \alpha(V) T
\end{equation}
and both
$\psi_2$ and $\phi_2$ are {\it homogeneous} functions
(of the same order) of the argument.
Supposing for a moment that $\phi_2 \propto {\cal V}^k $
is a homogeneous of the certain order $k$
[for the relativistic ideal gas
$\phi_2({\cal V})= N k_B/{\cal V}$, so $k=-1$]
one obtains
\begin{equation}
\psi_2 = (\gamma^{k+2}/\alpha) \phi_2.
\end{equation}
The resulting by the use of Eq.~(8) second equality
\begin{equation}
(\alpha^2/\gamma) \psi_1(\alpha T) = \phi_1(T),
\end{equation}
because $\phi_1$ for the relativistic ideal gas {\it explicitly}
appears to be {\it nonhomogeneous} function, can only be satisfied
if {\it identically}
\begin{equation}
\alpha(V) \equiv 1,
\end{equation}
and
\begin{equation}
\psi_1 = \gamma \phi_1.
\end{equation}
Thus the Lorentz  invariant temperature
is obtained as a {\it necessary
condition} for the
selfconsistence of thermodynamics.

Simultaneously, the entropy transforms like the energy,
\begin{equation}
dS^\prime= \gamma \phi_1 dT^\prime + \gamma^{k+2} \phi_2 d{\cal V}^\prime
\end{equation}
and $k=-1$,  $\phi_2({\cal V^\prime})= N k_B/{\cal V}^\prime$,
are obtained as a {\it necessary
and sufficient} condition for the
selfconsistence of thermodynamics.

Consequently the Gibbs--Boltzmann {\it thermodynamics}
of the ideal relativistic gas is {\it consistent},
and
\begin{equation}
T^\prime=T, \,
S^\prime= \gamma S, \,
{\cal P}^\prime =  \gamma^2 {\cal P},\,
U^\prime =  \gamma U
\end{equation}
are the resulting Lorentz transformation formulas.

The Eqs.~(14) for the ideal gas system are obtained rigorously,
just from the basic definition (1). They significantly differs
from the previous ones, where entropy and pressure have been
considered to be invariant, from
so-called Einstein--Planck formulas [2],
\begin{equation}
T^\prime=T/\gamma, \,
S^\prime= S, \,
{\cal P}^\prime = {\cal P},\,
U^\prime = U/\gamma
\end{equation}
and from the Ott formulas [3]
\begin{eqnarray}
U^\prime = \gamma U,\,  T^\prime=\gamma T,  \, S^\prime= S.
\end{eqnarray}
The different previous results are collected in Ref.~[1].
There is a number of reasons, for which those results should
be treated with care:

I) using the same {\it formalism} different groups have
obtained the different results and no consense has been reached;

II) the transformational properties are obtained
under the assumption that  the relations between
thermodynamical functions
and the canonical partition function
remain unchanged in moving reference frame;

III) the approach is tested on the ideal gas model, however the
incorrectly transformed partition function is used.

\noindent
There is the difference between transformational
properties of the probability density distributions of the ideal and
of the interacting system [4,5]. For the ideal gas the effect
of Lorentz transformation on space variables is counted
{\it only} globally ${\cal V}^\prime ={\cal V}/\gamma$ and thus the
momentum distribution change on non-covariant manner
\begin{equation}
\varepsilon W(\vec{p}) = \varepsilon^\prime W^\prime ({\vec p}^\prime),
\end{equation}
in order to conserve the probability measure. For interacting
particles there is no variable separation and the joint probability
density distribution in phase-space is invariant
\begin{equation}
f^\prime (\vec{x}^\prime_i,\, \vec{p}^\prime_i) = f(\vec{x}_i,\, \vec{p}_i).
\end{equation}
There is no contradiction between (17) and (18).
The special relativity requires
the full dynamical description of the system to be covariant. For the
projected kinetic, which is involved with the particular choice of
variables going to be reduced the Lorentz covariance is not already expected.
It is the reason that the first (non-covariant and true) rule can be
obtained from the general one. Both, (17) and resulting from the improper
use of (18) $f^\prime (\vec{p}^\prime) = f(\vec{p})$ cannot be
simultaneously true. It is the basic error of the previous consideration,
which manifests if the ideal gas is studied [1].

The particular {\it explicit} expressions are
presented for 1D case ($c=1$ and $N=1$ is used).
One has
\begin{equation}
W^\prime (p^\prime ) = \! { 1+V p^\prime /\sqrt{{p^\prime}^2 +m^2} \over
2m\sqrt{1-V^2} K_1(m\beta)}
\exp\Bigl[-{\sqrt{{p^\prime}^2 +m^2} + V p^\prime
\over \beta^{-1} \sqrt{1-V^2}} \Bigr],
\end{equation}
\begin{equation}
\langle \varepsilon^\prime \rangle = \int dp^\prime W^\prime (p^\prime )\sqrt{{p^\prime}^2 +m^2}=
{- m K_1^\prime (m\beta)  \over \sqrt{1-V^2}K_1 (m\beta)},
\end{equation}
\begin{equation}
\langle p^\prime \rangle =
{mV K_1^\prime (m\beta)  \over \sqrt{1-V^2} K_1 (m\beta)},
\end{equation}
\begin{equation}
Z_1^\prime ({\cal V}^\prime, \beta)= h^{-1} \sqrt{1-V^2} {\cal V}^\prime 2m K_1(m\beta).
\end{equation}
and, in the resting reference frame ($V=0$),
\begin{equation}
{\cal P} = \langle p^2/({\cal V}\sqrt{p^2+m^2}) \rangle =
1/\beta {\cal V} = k_B T /{\cal V}.
\end{equation}
\begin{eqnarray}
&S&=-k_B \langle \log [h W(p)/{\cal V}] \rangle  \nonumber \\ && =
k_B \{\beta \langle \varepsilon \rangle + \log[2m {\cal V}
K_1(m\beta)/h]\},
\end{eqnarray}
where the Planck constant is introduced for dimensional purposes and
$K_1$ is a modified Bessel function.

From Eqs.~(20--24) the general conclusion follows that
the {\it functional} relations
of the ordinary thermodynamics (in the resting reference frame)
are not invariant.
For instance, see (23),
$\gamma^2 {\cal P}={\cal P}^\prime
\ne k_B T^\prime/{\cal V}^\prime = \gamma {\cal P}$
Similarly, $-\partial  \ln (Z_1^\prime)/\partial \beta =
\langle \gamma [\varepsilon^\prime + p^\prime V] \rangle =
\langle \varepsilon \rangle$ do not correspond to the mean
energy (20) in the moving reference frame.

The important results of the paper are the following.
It has been shown that the previous transformational
formulas were incorrect. The most popular Einstein-Planck
relations (15) are consistent with the condition (i),
however they adopt incorrect energy transformation rule.
The Eqs.~(2) cannot be questioned, because except the trivial
statement that the average is a linear operation, these are
in fact
the definitions of the successive terms. The previously used
invariant properties of the entropy and the pressure are not true.
The pressure transformation formula (4), mentioned in Ref.~[6], has
been proven also by mechanical consideration. It has been shown
that the temperature is invariant, as was first suggested by
Landsberg [7]. The covariance of Eq.~(1) for ideal systems do not
require particular condition from $\phi_1 (T)$, so the relativistic
thermodynamics (1) of ideal gas is possible also
for the quantum statistics. The black body radiation
thermodynamics is not
of the form (3), but otherwise has the property that {\it all}
functions are {\it homogeneous}. This explains why
the consideration of the gas of photons cannot lead to
unambiguous conclusions. In fact, because $U=c{\cal V} T^4$,
${\cal P} = cT^4/3$, and $S=4c {\cal V} T^3/3$, then the assumption
of arbitrary transformational rules of the form
\begin{eqnarray}
&&{\cal V}^\prime= {\cal V}/\gamma, \, T^\prime=\gamma^\tau T, \nonumber \\
&&S^\prime({\cal V}^\prime, \,T^\prime)  = \gamma^\sigma S({\cal V},\, T) =
 \gamma^{\sigma+1-3\tau}  4c{\cal V}^\prime {T^\prime}^3/3, \nonumber \\
&& {\cal P}^\prime =  \gamma^\pi {\cal P}= \gamma^{\pi-4\tau} c{T^\prime}^4/3, \nonumber \\
&&U^\prime =  \gamma^u U= \gamma^{u+1-4\tau} c {\cal V}^\prime {T^\prime}^4,
\end{eqnarray}
results with only two equations for the four variables
\begin{equation}
u+1 = \pi, \quad u=\sigma+\tau,
\end{equation}
in order to retain the covariance of Eq.~(1).
Two arbitrary assumptions, $\pi=0$ and $\sigma=0$, leading to Eqs.~(15)
were incorrect. The proper choice, $u=1$ results with $\pi=2$, however
$\tau$ and $\sigma=1-\tau$ remain undetermined.
The consideration of ideal gas results with $\tau=0$ and $\sigma=1$.

\end{document}